\documentclass[%
 reprint,
superscriptaddress,
 amsmath,amssymb,
 aps,
]{revtex4-2}
\usepackage{graphicx}% Include figure files
\usepackage{dcolumn}% Align table columns on decimal point
\usepackage{bm}% bold math
\usepackage{xcolor}

\begin{document}

\preprint{APS/123-QED}

\title{Electromagnon signatures of a metastable multiferroic state}% Force line breaks with \\

\newcommand{\MITCHEM}{Department of Chemistry, Massachusetts Institute of Technology, Cambridge, Massachusetts 02139, USA}
\newcommand{\ORNL}{Neutron Scattering Division, Oak Ridge National Laboratory, Oak Ridge, Tennessee 37831, USA}
\newcommand{\AUSPHYS}{Department of Physics, The University of Texas at Austin, Austin, Texas 78712, USA}

\newcommand{\Chongqing}{Center of Quantum Materials and Devices, Chongqing University, Chongqing 401331, China}

\author{Blake S. Dastrup}
\altaffiliation{These authors contributed equally to this work}
\affiliation{\MITCHEM}
\author{Zhuquan Zhang}
\altaffiliation{These authors contributed equally to this work}
\affiliation{\MITCHEM}
\author{Peter R. Miedaner}
\affiliation{\MITCHEM}
\author{Yu-Che Chien}
\affiliation{\MITCHEM}
\author{Young Sun}
\affiliation{\Chongqing}
\author{Yan Wu}
\affiliation{\ORNL}
\author{Huibo Cao}
\affiliation{\ORNL}
\author{Edoardo Baldini}
\email{edoardo.baldini@austin.utexas.edu}
\affiliation{\AUSPHYS}
\author{Keith A. Nelson}
\email{kanelson@mit.edu}
\affiliation{\MITCHEM}

\begin{abstract}
Magnetoelectric multiferroic materials, particularly type-II multiferroics where ferroelectric polarizations arise from magnetic order, offer significant potential for the simultaneous control of magnetic and electric properties. However, it remains an open question as to how the multiferroic ground states are stabilized on the free-energy landscape in the presence of intricate competition between the magnetoelectric coupling and thermal fluctuations. In this work, by using terahertz time-domain spectroscopy in combination with an applied magnetic field, photoexcitation, and single-shot detection, we reveal the spectroscopic signatures of a magnetic-field-induced metastable multiferroic state in a hexaferrite. This state remains robust until thermal influences cause the sample to revert to the original paraelectric state. Our findings shed light on the emergence of metastable multiferroicity and its interplay with thermal dynamics.
\end{abstract}

%\keywords{Suggested keywords}%Use showkeys class option if keyword
                              %display desired
\maketitle

%\tableofcontents

% \section{\label{sec:Intro}Introduction}
Magnetoelectric (ME) multiferroics, characterized by the coexistence and coupling of ferroelectric and magnetic orders, have attracted widespread interest due to their emergent physical properties and potential applications in next-generation electronics and information technologies \cite{eerenstein2006multiferroic,spaldin2010multiferroics,fiebig2016evolution,spaldin2019advances,pimenov2006possible,tokura2014multiferroics,matsukura2015control,dong2019magnetoelectricity}. Despite the identification of many multiferroic materials with strong magnetoelectric coupling \cite{heyer2006new,kimura2003magnetic,kimura2008cupric,song2022evidence,gao2024giant}, very few of them exhibit magnetoelectric effects at room temperature \cite{kitagawa2010low,valencia2011interface,chun2012electric}. Stabilizing multiferroic states at elevated temperatures therefore remains a long-standing challenge \cite{roy2012multiferroic,kimura2012magnetoelectric, vopson2015fundamentals}. Key to overcoming this obstacle is a comprehensive understanding of the underlying mechanisms governing multiferroic order and the delicate balance between various interactions that either favor or hinder multiferroic states. In this context, metastable multiferroic states, distinguished by their long-lasting nature and resilience to thermal perturbations \cite{pimenov2006possible,lee2012heliconical,nakajima2016electromagnon}, offer a unique platform for investigating the complex interplay between ME coupling and thermodynamic fluctuations. Systematic studies of such metastable states can provide fundamental mechanistic insights into the stability of multiferroic phases, ultimately facilitating the design and synthesis of novel materials with multiferroic behavior at high temperatures. Although some evidence suggests the existence of metastable states in certain multiferroics \cite{lee2012heliconical,nakajima2016electromagnon,zhai2017giant,kocsis2020stability}, the processes through which these states emerge and stabilize within complex energy landscapes—where competing interactions and parameter fluctuations are at play—remains elusive.

In this study, we focus on the Y-type hexaferrite Ba$_{2-x}$Sr$_x$Mg$_2$Fe$_{12}$O$_{22}$ (BSMFO) with $x=1.6$ \cite{zhai2017giant} as a model system. As shown in Fig. 1(a), the crystal structure of BSMFO belongs to the space group $R\bar{3}m$ with alternating stacked subunits of S (Mg$_2$Fe$_4$O$_8$) and T ((Ba,Sr)$_2$Fe$_8$O$_{14}$) blocks. The magnetic moments of individual Fe ions are arranged roughly collinearly, forming separate sets of magnetic subunits with large (L) and small (S) magnetic moments. These renormalized magnetic moments order in an incommensurate longitudinal conical (LC) structure along the crystallographic [001] axis below $\sim$120 K (see Fig. 1(b)). Applying an external DC magnetic field perpendicular to the [001] axis beyond a certain threshold can switch the magnetic state to a transverse conical (TC) phase with either four-fold or two-fold symmetry. While the LC state is paraelectric, the TC state exhibits ferroelectric order resulting from the spin current or the exchange striction \cite{katsura2005spin,mostovoy2006ferroelectricity,kida2011gigantic}. It has been demonstrated that the specific level of Sr doping with $x=1.6$ significantly enhances the ME coefficient by suppressing the four-fold symmetric TC state and lowering the energy barrier for the ferroelectric phase transition to the two-fold spin state \cite{zhai2017giant}, as depicted in Fig. 1(c). This Sr-doped variant of Y-type hexaferrite displays unconventional hysteresis upon transitioning to the two-fold TC state \cite{zhai2017giant}, making it a promising candidate for the investigation of metastable multiferroic states.

Here, we use terahertz time-domain magnetospectroscopy (THz-TDMS) to probe a magnetic collective mode, specifically the electromagnon excitation that characterizes different magnetic phases in this composition-tuned Y-type hexaferrite. Unlike conventional magnetic resonances \cite{kampfrath2011coherent,zhang2024terahertz,zhang2024coupling}, these low-energy excitations originate from dynamical ME coupling \cite{katsura2007dynamical,takahashi2012magnetoelectric,takahashi2013terahertz} and can directly couple to the electric field of a THz beam with large dipole moments, thereby offering a unique fingerprint of the intrinsic magnetic structures \cite{kida2009electric,ishiwata2010neutron,kida2011gigantic,nakajima2016electromagnon,chun2018electromagnon,ueda2023non}. By monitoring changes to such collective mode as a function of magnetic field and temperature, we unambiguously identify the metastability of the switched ferroelectric state and its competition with thermal fluctuations.

\begin{figure}[b]
\centering
\includegraphics[width = 85 mm]{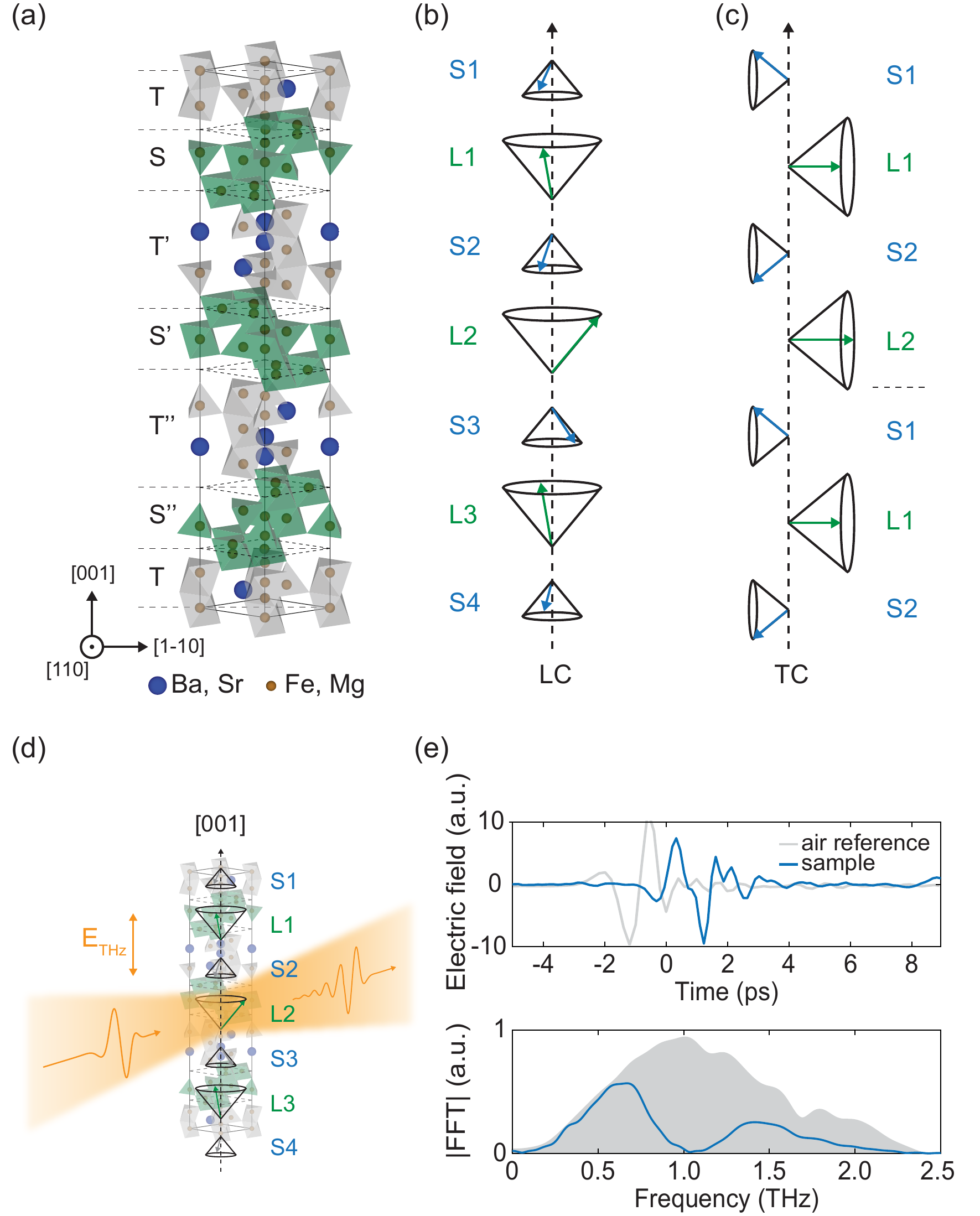}
\caption{\label{fig:1} Crystal and magnetic structures of BSMFO. (a) Crystal structure of BSMFO with structural subunit boundaries denoted on the left and magnetic subunits in the LC and TC phases denoted on the right. Below 120 K a conical spin state forms with cones oriented along the [001] axis (LC) as shown in (b). In the presence of a transverse B-field, the cone orientation becomes perpendicular to the [001] axis (TC) as shown in (c). (d) Schematic illustration of the measurement configuration with the THz electric field oriented along the [001] axis. (e) An example THz time-domain trace at B = 0 T (top panel) and THz amplitude spectrum (bottom panel) obtained by Fourier transformation of the time-domain data.}
\end{figure}

% \section{\label{sec:Results}Results}
As an initial step, we established the detection of the electromagnon mode in the THz frequency range using THz-TDMS. In our experiment, we focused a nearly single-cycle THz pulse generated through optical rectification into either the sample or air, and we measured the corresponding time-domain transmission signal or reference via electro-optic sampling. Upon cooling the sample to 5 K without an externally applied magnetic field, the system stabilizes in the LC state, and the Fourier transform of the THz transmission signal exhibits a pronounced broad dip at 1.1 THz compared to the reference spectrum, as seen in Fig. 1(e). This feature is observed only when the THz electric field is aligned along the [001] axis (i.e., $\mathbf{E}_{THz} \parallel [001]]$, see Fig. 1(d)), which is consistent with previous reports \cite{kida2009electric,kida2011gigantic,nakajima2016electromagnon} and the selection rule of the electromagnon mode in these helimagnetic systems. Throughout this paper, we shall use the term ``electromagnon'' to refer broadly to electric-dipole-active magnetic excitations in both paraelectric and ferroelectric phases, in line with the literature \cite{kida2009electric,kida2011gigantic,nakajima2016electromagnon,chun2018electromagnon}, rather than limiting it to the narrower definition of an emergent excitation from multiferroic ordering.

To validate our identification of this broad absorption with the electromagnon mode, we measured the magnetic field dependence of the THz spectrum. To this aim, we initially aligned the external magnetic field parallel to the [001] axis. Then, we recorded the THz spectrum, defined as $A(\Omega)=1-\left |T(\Omega) \right |$, where $T(\Omega)=\tilde{E}_{sample}(\Omega)/\tilde{E}_{ref}(\Omega)$ is the THz transmittance, as a function of external magnetic field $B$ (see Fig. 2(a)). From $B = 0$-$0.2$ T, we observe an initial narrowing of the resonance, which we ascribe to magnetic domain alignment. After this initial narrowing, the resonance frequency exhibits a blueshift from $B = 0.2 - 1.1$ T, followed by a redshift from $B = 1.1 - 6.2$ T. In this case, increasing the magnetic field reduces the conical angle $\theta$ between the magnetic moments and the [001] axis. Beyond $B = 6.2$ T, the absorptive signal is no longer discernible, indicating that the sample has entered a magnetic-field induced longitudinal ferrimagnetic (spin-collinear) phase, wherein the magnetic moments L and S align antiparallel along the [001] axis. Based on the exchange-striction excitation model \cite{kida2011gigantic,chun2018electromagnon,zhai2018electric}, no electromagnon activity is anticipated for conical angles at both extremes, $\theta$ = 0° or 90°, which naturally accounts for the observation that the electromagnon reaches its maximum frequency at an intermediate conical angle and then vanishes in the ferrimagnetic phase.

\begin{figure}[b]
\centering
\includegraphics[width = 85 mm]{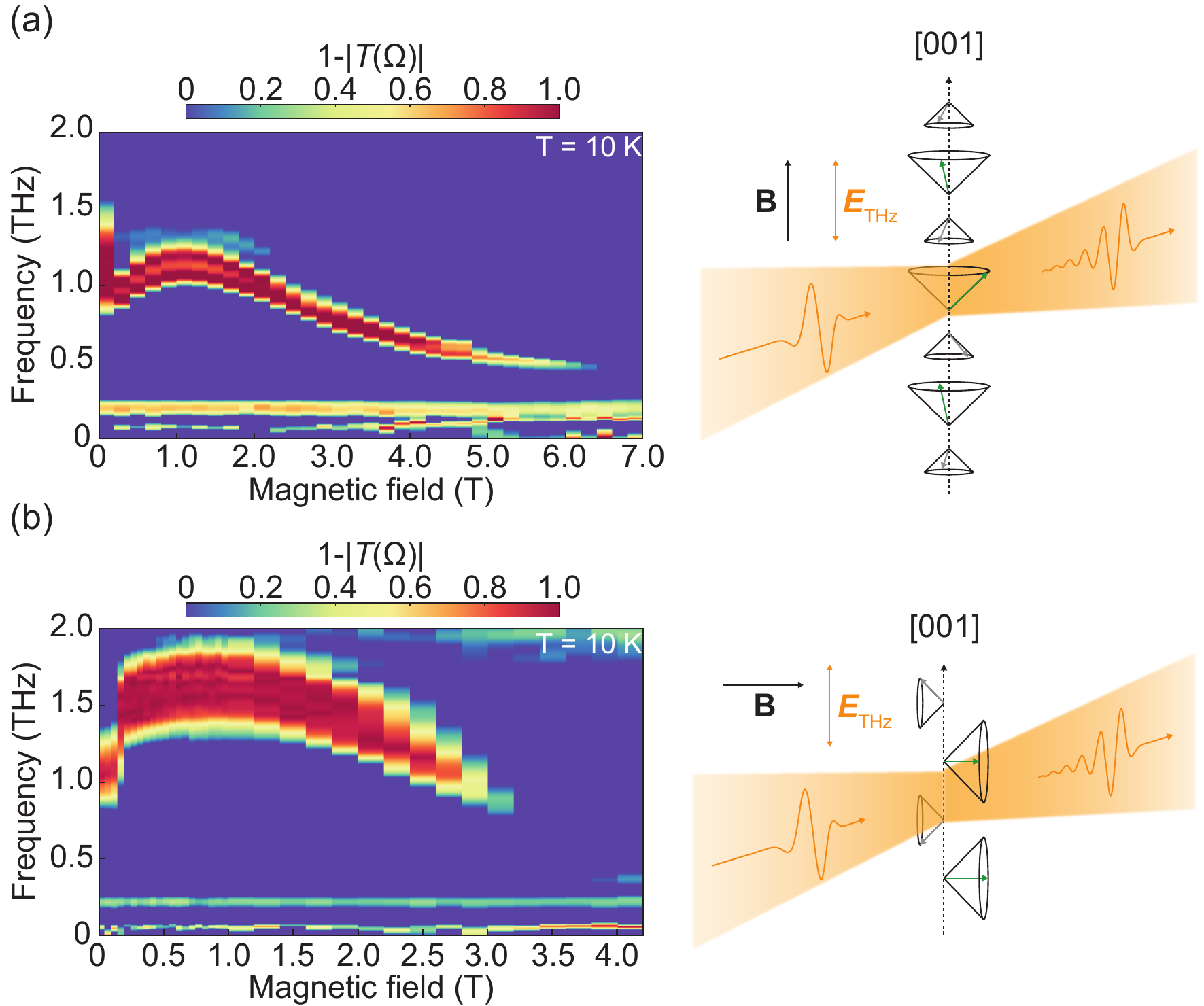}
\caption{\label{fig:2} Magnetic field dependence of the electromagnon at 10 K. The THz spectrum ($1-|T(\Omega)|$) is plotted as a function of the external magnetic field $\mathbf{B}$. (a) $\mathbf{B} \parallel$ [001]. The shift in frequency is caused by a declination of the cone angle toward the [001] axis until a ferrimagnetic state is reached with spin moments aligned along [001]. (b) $\mathbf{B} \parallel$ [100]. The rapid shift at B = 0.12 T is caused by a magnetic-field-induced transition from the LC to the TC state. Further shifting of the frequency after the switch is due to the declination of the cone angle perpendicular to [001] until a ferrimagnetic state is reached with spin oriented perpendicular to [001].}
\end{figure}

Next, we oriented the magnetic field to be perpendicular to the [001] axis (i.e., $\mathbf{B} \parallel [100]$) to switch the system into the ferroelectric TC state. In this configuration, the THz spectrum shows negligible change until $B=0.12$ T, at which point a discontinuous shift in the electromagnon frequency from 1.1 THz to 1.6 THz is observed, signifying a first-order phase transition. This abrupt change in the electromagnon frequency contrasts with the continuous phase transitions observed in similar compounds, marking it as a distinctive fingerprint of the ferroelectric TC state in this material. Upon further increasing the magnetic field, the electromagnon resonance redshifts until it is no longer discernable above 3.2 T. Similar to the previous scenario, this behavior corresponds to the reduction in the conical angle associated with the TC state, which ultimately evolves into the transverse ferrimagnetic state at high magnetic fields, with L and S magnetic moments perpendicular to the [001] axis. This frequency shift of the mode can be well captured by modeling the system with an effective four-spin Hamiltonian \cite{chun2018electromagnon} (see Supplementary Materials). Notably, upon transitioning the sample to the TC state and then setting the magnetic field to zero, the electromagnon frequency remains at the shifted value of 1.5 THz. This behavior suggests that, below a certain temperature, the TC state is metastable.

\begin{figure}[b]
\centering
\includegraphics[width = 85 mm]{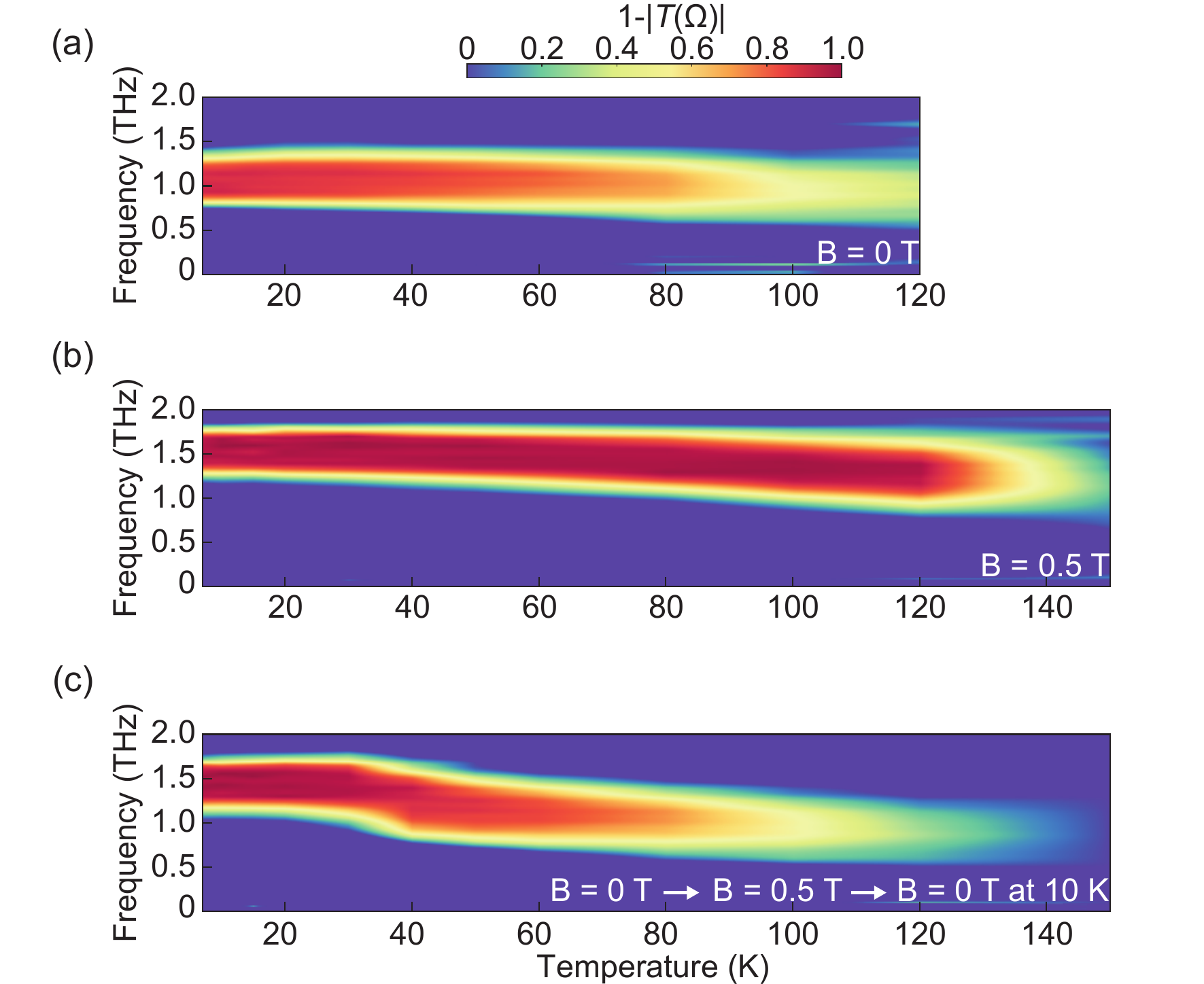}
\caption{\label{fig:3} Temperature dependence of the electromagnon spectrum in the LC and TC states. THz spectrum (1-$|T(\Omega)|$) plotted as a function of temperature for (a) initial cool-down at B = 0 T, (b) temperature ramp with B = 0.5 T, and (c) temperature ramp following a zero-field cooldown and subsequent cycling of the magnetic field from B = 0.5 T back to B = 0 T to induce the metastable TC state. The data in (a) reveal the electromagnon resonance starting at ~120 K. Similarly, the electromagnon signal in the TC state, shown in (b), persists under external magnetic field up to ~140 K without switching back to LC. If no field is present, as in (c), the TC state is metastable until ~40 K, above which the electromagnon frequency reverts back to the value observed during zero-field cooling. Note that the plots as shown have been interpolated for ease of viewing. See Supplementary Materials for raw data.}
\end{figure}

To further explore the metastability of the multiferroic TC state, we performed a set of measurements in which the electromagnon signal was probed by THz-TDMS over a broad range of temperatures. The results are shown in Fig. 3. Starting from 120 K, we cooled the sample without applying an external magnetic field and measured the THz spectrum at each temperature. As shown in Fig. 3(a), the electromagnon signal is already present at approximately 120 K and grows in strength as the temperature is lowered to a minimum of 10 K. Despite an initial hardening of the mode during the cool-down (i.e., above 80 K), the center frequency of the electromagnon signal is close to 1.1 THz, indicating that the system stabilizes in the LC state. Once the sample temperature reached 10 K, we applied a 0.5 T magnetic field perpendicular to the [001] axis, well above the switching threshold, to transition the LC state to the TC state. With the field applied, the electromagnon signal appears at the shifted frequency, i.e., 1.5 THz (see Fig. 3(b)). We then gradually increased the temperature and monitored the THz spectrum while maintaining the magnetic field. We observe that although the electromagnon mode continuously softens and its linewidth broadens with increasing temperature, its frequency never reaches 1.1 THz, and the signal eventually disappears near 140 K. This result suggests that such a magnetic field can induce ferroelectricity over a wide range of temperatures where the system starts in the LC state. Lastly, we set the magnetic field to 0 T and lowered the temperature back to 10 K, at which point the sample was again switched into the TC state by cycling the magnetic field to 0.5 T and then setting it to 0 T afterward. After cycling the field, the electromagnon mode appears at the TC frequency (see Fig. 3(c)). The signal persists at this frequency until the temperature reaches ~30 K, where the mode begins to soften. Above 40 K, the spectrum matches the signal observed during the initial zero-field cooling, indicating that the sample has reverted to the LC state.
\begin{figure}[!t]
\centering
\includegraphics[width = 85 mm]{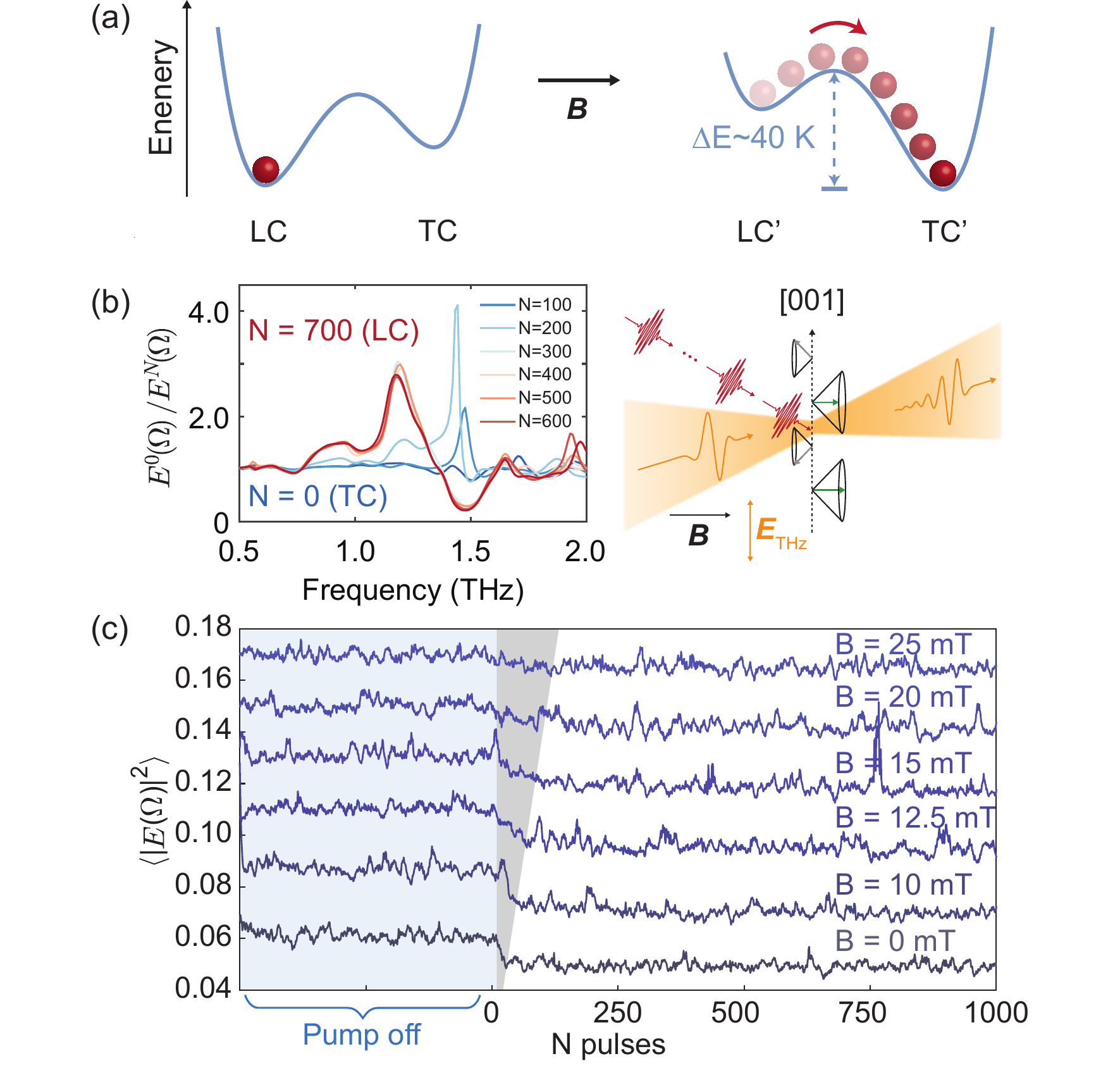}
\caption{\label{fig:4} Thermal switching dynamics measured by single-shot THz detection. (a) Simplified illustration of the low-temperature magnetic-field-dependent free energy landscape of BSMFO. (b) Left: THz ratio spectra ($\tilde{E}^0(\Omega)/\tilde{E}^N(\Omega)$) are shown for different numbers of pump pulse accumulations, N, in increments of 100. (See Supplementary Materials Fig. S5 for the individual raw spectra $\tilde{E}^N(\Omega)$.) At N = 0, the spectrum matches the TC state, but after 700 shots, the spectrum has progressed to that of the LC state. Right: Schematic illustration of the measurement configuration with THz electric field along the [001] axis, in combination with laser heating. (c) The integral of the THz spectrum is plotted as a function of time, where t = 0 marks the opening of the pump pulse shutter. Curves are shown for B = {0, 10, 12.5 15, 20, 25} mT. As the magnetic field increases, the slope of the curve becomes less steep, indicating that more energy must be accumulated for switching to occur (i.e., well-deepening).}
\end{figure}

From the above measurements, we can conclude that after switching, the multiferroic TC state is metastable below 30 K. The temperature-sensitivity of the observed metastability further suggests that thermal fluctuations play a major role in switching the system back to the paraelectric state (see Fig. 4(a)). As the temperature increases above 40 K, the thermal energy (i.e., $\sim$3.4 meV) becomes sufficient to overcome the energy barrier between the TC and LC states, leading to the destabilization of the TC state. To better understand the competition between the thermal fluctuations and magnetic ordering in determining the free energy landscape of the system, we set the temperature of the sample to 30 K near the temperature threshold where natural relaxation to the LC state occurs, and then exposed the sample to a sequence of laser pulses separated by 1 ms with a pump photon energy at 1.55 eV and an incident fluence of 7.08 mJ/cm$^2$ to thermally perturb the sample. For clarity, this measurement is not a time-resolved measurement of ultrafast (e.g. femtosecond to picosecond) dynamics, but rather a single-shot measurement of transmitted THz signals \cite{gao2022snapshots,gao2022high,dastrup2024optical} to probe the system as it evolves thermally on the timescale of milliseconds (see Supplementary Materials). Figure. 4(b) shows the evolution of the ratio between the transmission spectra before and after photothermal excitation, defined as $\tilde{E}^0(\Omega)/\tilde{E}^N(\Omega)$, where N is the number of accumulated laser pulses and $\tilde{E}^N(\Omega)$ is the Fourier transform of the THz transmission after exposure to N laser shots. In this context, a signal smaller (larger) than unity indicates that a certain spectral component has lost (gained) its spectral weight, making the material less (more) transparent. The data reveal that after exposure to the laser pulses, the electromagnon signal at 1.5 THz first undergoes a redshift in frequency, as evidenced by the dispersive feature at 1.5 THz, and then diminishes, transferring its spectral weight to a broad resonance around 1.1 THz. This spectral feature persists even after the laser pulses are blocked. This behavior indicates the system progresses from the ferroelectric TC state to the paraelectric LC state after cumulative laser heating in the absence of a magnetic field. The pump pulse irradiates the front surface of the sample and deposits energy into the excited region defined by the beam spot size and optical penetration depth. Thermal transport from there induces a portion of the material larger than the excited volume to switch to the LC state. As a result, it leads to a partial restoration of the LC spectrum, while some material remains in the TC state, causing a decrease but not complete elimination of the TC spectrum. We repeated this measurement for different values of the magnetic field ($\mathbf{B} \parallel [100]$) below the switching threshold. This allowed us to more sensitively track the interplay between these two thermodynamic perturbations—magnetic field and temperature—by monitoring the number of pulses required to reach a thermally steady state. Figure 4(c) shows integrated THz spectral amplitude as a function of time before and during exposure to laser pulses for the different magnetic field values. At zero field, we see an abrupt change in the value of the integral where the sample goes from the TC state back to the LC state. As the magnetic field increases, this transition time becomes much slower and the level of transition gets weaker, indicating that more thermal energy must be accumulated to reach the threshold where the system can relax back to the LC state. At 25 mT, the laser-induced heating is able to switch the sample back to the LC state, but only for a shallower depth within the sample, explaining the smaller ultimate change in the THz spectrum. These findings suggest that an increasing magnetic field raises the energy barrier to restore the LC state, thus stabilizing the metastable TC phase at these temperatures.

% \section{\label{sec:Conclusion}Conclusion}
Taken together, our THz-TDMS measurements elucidate the electromagnon signatures that correspond to the paraelectric and multiferroic phases of the Y-type hexaferrite Ba$_{0.4}$Sr$_{1.6}$Mg$_2$Fe$_{12}$O$_{22}$. By tracing the evolution of the electromagnon mode, coupled with the insights garnered from the magnetic field, temperature, and laser shot dependence measurements, we provide evidence for a metastable multiferroic state induced by a moderate magnetic field up to 30 K.  Our results also underscore the intricate balance between thermal fluctuations and the magnetic anisotropy barrier in facilitating the stability of this metastable state. This discovery is potentially informative in the pursuit of high-temperature multiferroic materials, which are of paramount importance in the development of emerging energy and information technology applications. Our findings suggest approaches to the design of innovative materials that are capable of withstanding higher temperatures while preserving their multiferroic properties. Moreover, our experimental methodology presents an advanced platform that enables the discovery of other exotic phases inherent in magnetic quantum materials.

\begin{acknowledgments}
The MIT researchers were supported by the U.S. Department of Energy, Office of Basic Energy Sciences, under Award No. DESC0019126. Y.S. was supported by the National Natural Science Foundation of China (Grant No. 12227806). E.B. was primarily supported by the Air Force Office of Scientific Research under Young Investigator Program award FA9550-24-1-0097. A portion of this research used resources at the High Flux Isotope Reactor, US Department of Science (DOE) Office of Science User Facilities operated by Oak Ridge National Laboratory (ORNL).
\end{acknowledgments}

\bibliography{main}% Produces the bibliography via BibTeX.

\end{document}